\documentclass[preprint]{aastex}

\usepackage{graphicx}
\usepackage{comment}
\usepackage{epstopdf}
\newcommand{\htcop}{H$^{13}$CO$^+$}
\newcommand{\kms}{km~s$^{-1}$}

\shorttitle{Magnetic Fields in IRAS~16293} \shortauthors{Rao et al.}

\begin{document}

\title{IRAS~16293: A ``Magnetic'' Tale of Two Cores}

\author{Ramprasad Rao\altaffilmark{1}}
\affil{Academia Sinica Institute of Astronomy and Astrophysics, Taipei, Taiwan}
\email{rrao@sma.hawaii.edu}

\author{Josep M. Girart}
\affil{Institut de Ci\`encies de l' Espai (CSIC-IEEC), Campus UAB
-- Facultat de Ci\`encies, Torre C5 - parell 2, 08193 Bellaterra,
Catalunya, Spain}

\author{Daniel P. Marrone\altaffilmark{2}}
\affil{National Radio Astronomy Observatory \& Kavli Institute for
Cosmological Physics, University of Chicago, 5640 South Ellis Avenue,
Chicago, Illinois, 60637, USA}

\author{Shih-Ping Lai}
\affil{Institute of Astronomy and Department of Physics, National
Tsing-Hua University, Taiwan}

\and

\author{Scott Schnee}
\affil{NRC-Herzberg Institute of Astrophysics, 5071 W. Saanich Road,
Victoria, BC, V9E 2E7, Canada}

\altaffiltext{1}{Mailing Address: Submillimeter Array, 645 N. Aohoku Pl, Hilo, HI 96720, USA}
\altaffiltext{2}{Jansky Fellow}

\begin{abstract} 

We present  polarization observations of the dust continuum emission
from the young star forming region IRAS~16293. These observations
of IRAS~16293, which is a binary system, were conducted by the
Submillimeter Array (SMA) at an observing frequency of 341.5 GHz
($\lambda \sim 880\mu$m) and with high angular resolution
($\sim$2$\arcsec$--3$\arcsec$).  We find that the large scale global
direction of the field, which is perpendicular to the observed
polarization, appears to be along the dust ridge where the emission
peaks.   On smaller scales we find that the field structure is
significantly different for the two components of the binary. The
first component, source A, shows a magnetic field structure which
is ``hourglass'' shaped as predicted from theoretical models of low
mass star formation in the presence of strong magnetic fields.
However, the other component, source B, shows a relatively ordered
magnetic field with no evidence of any deformation.  We have possibly
detected a third younger outflow from source A as seen in the SiO
emission which is in addition to the two well known powerful bipolar
outflows in this kinematically active region.  There is an observed
decrease in polarization towards the center and this ``polarization
hole'' is similar to decreases seen in other young star forming
regions.  Our calculations show that in IRAS~16293 the magnetic
energy is stronger than the turbulent energy but is approximately
similar to the centrifugal energy. There is considerable misalignment
between the outflow direction and the magnetic field axis and this
is roughly in agreement with model predictions where the magnetic
energy is comparable to the centrifugal energy. In conjunction with
other observations of the kinematics  as determined from the outflow
energetics and chemical differentiation we find that our results
provide additional evidence to show that the two protostars appear
to be in different stages during their evolution.

\end{abstract}


\section{Introduction}

In the ``classical picture'' of star formation, magnetic fields are
believed to strongly influence star formation activity in molecular
clouds. They provide support to a cloud against gravitational
collapse and thus explain the low efficiency of the star formation
process \citep{Mouschovias01, Shu07}. The process of ambipolar
diffusion, in which magnetic flux is redistributed in the cloud,
leads to the formation of a core that can no longer be magnetically
supported and gravitational collapse sets in. In addition, the
process of magnetic braking can help to remove angular momentum and
slow down the rotation of the cloud as it collapses \citep{Basu94}.
In contrast, there have been a number of alternate theories which
postulate that magnetic fields are relatively weak and supersonic
magnetohydrodynamic turbulence is the dominant process \citep{MacLow04}.
Turbulence controls the evolutions of clouds, and cores form at the
intersection of supersonic turbulent flows. Only a fraction of such
cores become supercritical and collapse begins to occur on a
gravitational free-fall timescale \citep{Ballesteros07}.

Studying the morphology of the magnetic field in the interstellar
medium (ISM) requires difficult and sensitive observations of
polarized radiation \citep{Hildebrand04}.  Spinning dust grains in
the ISM become partially aligned with the magnetic field, generally
with their long axes perpendicular to the field \citep{Davis51}.
Nevertheless, the exact nature of the alignment process is a matter
of debate \citep{Lazarian07}, but it appears that the dominant
mechanism may actually be alignment by radiative torques \citep{Hoang08,
Hoang09}.  As a consequence of this alignment, the thermal dust
emission is partially linearly polarized, with polarization direction
perpendicular to the magnetic field.  The magnitude of the field
can be determined indirectly by using the dispersion of the position
angles of the field under the assumption of energy equipartition
between the kinetic and perturbed magnetic energies \citep{Chandra53,
Ostriker01,Heitsch01,Crutcher04,Falceta08}.

Observations of the large scale polarization distribution in molecular
clouds  from optical, infra-red, and submillimeter (submm) wavelengths
usually show an ordered pattern \citep{Dotson00, Pereyra04, Poidevin06,
Matthews09}. The analysis of polarization data has showed that some
dark clouds are  magnetically dominated \citep{Alves08, Heyer08}
while other giant molecular clouds are close to equipartition with
turbulence \citep{Novak09}.  However, observationally, the debate
on whether turbulence or magnetic fields are dominant is still
unresolved \citep{Crutcher09}.  Early and pioneering work with the
BIMA millimeter~(mm) array showed the dust (and molecular line)
polarized emission can be traced at an angular resolution of few
arcseconds, the scales where the core contraction process occurs
\citep{Rao98,Girart99,Lai01,Lai03,Cortes05}. Currently, the
Submillimeter Array (SMA) is the only telescope that can detected
the polarized emission at mm and submm wavelengths  \citep{Marrone06b,
Girart06, Girart09,Tang09a, Tang09b}. In the object NGC~1333~IRAS~4A
(IRAS~4A here after), the SMA observations have revealed that  the
magnetic field configuration is consistent with theoretical models
for the formation of solar-type stars in which magnetic fields play
a much stronger role than turbulence \citep{Girart06, Goncalves08}.

The low-mass star forming region IRAS~$16293-2422$ (IRAS~16293 here
after), located in the $\rho$ Ophiuchi molecular cloud complex, has
been the focus of numerous studies since the report of infall
spectral signatures \citep{Walker86,Menten87}.  Accurate observations
using astrometric techniques show that the $\rho$ Ophiuchi molecular
cloud is at a distance $\simeq 120$~pc \citep{Knude98,Loinard08},
while maser VLBI observations towards IRAS~16293 suggest a distance
of $\simeq 178$~pc \citep{Imai07}.  For this paper we have assumed
a distance of $d=150$~pc.  IRAS~16293 is a Class~0 protostellar
system with a bolometric luminosity of 32~L$_{\sun}$, and is
surrounded by a compact, $R_{\rm env} \simeq3000$~AU, but relatively
massive envelope, $\simeq3.0$~M$_{\sun}$, \citep{Correia04}. Higher
resolution radio interferometric observations have revealed a double
core separated by  5$\arcsec$ (750~AU in the plane of the sky). The
southern core is commonly referred to as source A and the northern
one as source B. Both cores appear to have different physical and
chemical properties \citep{Wootten89, Estalella91,
Bottinelli04,Chandler05,Takakuwa07}.  Despite its low luminosity,
IRAS~16293 has a rich chemistry, with hot-core like properties at
scales of $\sim100$~AU  \citep{Blake94, Ceccarelli00, Schoier02,
Cazaux03, Kuan04, Bisschop08}. There is a strong quadrupolar outflow
associated with source~A \citep{Walker88,Mizuno90,Stark04,Yeh08}.
The larger bipolar outflow is in the east-west direction (E-W
outflow), with a position angle (PA) of $110\arcdeg$. The other
outflow is in the northeast-southwest direction (NE-SW outflow),
with a PA of $60\arcdeg$ and shows copious SiO emission \citep{Hirano01}.
High resolution submm observations \citep{Chandler05} have revealed
multiplicity in source A and which of these sources powering these
two outflows is a matter of debate \citep{Loinard07}.  Previous
measurements of the magnetic field geometry in this object by
observing polarized dust emission have yielded contradictory results
\citep{Flett91,Tamura93,Akeson97}, possibly because they were
sensitivity limited.  Thus, it is apparent that more sensitive and
higher resolution observations of the magnetic field structure are
needed.

In \S~2 we briefly describe the observations and the data reduction
procedure. \S~3 presents the results, \S~4 the analysis, and \S~5
discusses the possible scenarios for the observed magnetic field
morphology in context with the information from observations already
known from the literature.

\section{Observations and Data Analysis}

The observations were conducted in April 2006 (see details in
Table~\ref{obspar}) with the SMA\footnote{The Submillimeter Array
is a joint project between the Smithsonian Astrophysical Observatory
and the Academia Sinica Institute of Astronomy and Astrophysics and
is funded by the Smithsonian Institution and the Academia Sinica},
which is located near the summit of Mauna Kea in Hawaii \citep{Ho04}.
As can be seen from the Caltech Submillimeter Observatory (CSO) tau
meter atmospheric opacities in Table~\ref{obspar}, the observing
conditions on the first day were significantly better than those
on the second day.  Nevertheless, the atmospheric conditions were
stable on both days resulting in minimal fluctuations in the antenna
gains.  The coordinates of the pointing center was at RA=$16^h32^m22\fs9$
and Dec.=$-24\arcdeg28\arcmin36\farcs00$. The observing frequency
was chosen to be located in the 345~GHz atmospheric window, which
at the Mauna Kea site provides both optimal sensitivity and angular
resolution. The SMA receivers operate in a double sideband mode
with the two sidebands separated by $\sim$10~GHz and the selected
local oscillator frequency placed the lower and upper sideband
central frequencies at 336.5~GHz and 346.5~GHz respectively. The
SMA correlator has a bandwidth of $\sim$2~GHz which comprises of
24 partially overlapping spectral windows. All the spectral windows
had 128 channels, which provided a velocity resolution of 0.7~\kms.
In addition to the continuum, the tuning frequency and the correlator
were chosen to  allow for simultaneous observation of the emission
from the CO~3--2, SiO~8--7, and H$^{13}$CO$^+$~4--3 spectral lines.

Conducting polarimetric observations with interferometer arrays at
mm and submm wavelengths is challenging and requires the use of
some special techniques.  A brief description of these techniques
is provided in \citet{Marrone06b} and a more detailed discussion
of the methodology (both hardware and software aspects) is available
in \citet{Marrone06a} and \citet{Marrone08}.  The data were reduced
using the MIRIAD software package \citep{Wright93}.  The instrumental
gains were calibrated by interspersing observations of IRAS~16293
(the target source) with observations of the quasars J$1517-243$
and J$1622-297$ which were used as gain calibrators.  The instrumental
spectral bandpass was calibrated from observations of the quasar
3C~273 and the absolute flux scale was determined from observations
of Callisto.  The single greatest factor that can corrupt the data
is the intrinsic instrumental polarization which is commonly referred
to as the ``leakage''.  Therefore, the data were needed to be
carefully calibrated to remove the effects due to this leakage. The
primary task in MIRIAD which was used to solve for the leakage is
GPCAL.  For our observations we used 3C~273, whose intensity and
polarization are strongly variable.  In addition to the leakage,
the task GPCAL also simultaneously solves for the polarization of
the calibrator as well.  At the epoch of our observation, we
determined the intensity to be $\sim$8.7~Jy and the linear polarization
to be $1.0\pm0.1$\% at a position angle of $-34\arcdeg$.  The
leakages are different in each of the two sidebands as they differ
in frequency by $\sim$10~GHz.  In the upper sideband, which is
closer to the design frequency of the SMA polarimetry system, the
measured leakages were approximately 1\%, while the lower sideband
leakages were between 2 and 3\%. These leakages were measured to
an accuracy of 0.2\% or better.

The data from the source of interest, IRAS~16293, were then corrected
for the leakages. The continuum emission in each of the two sidebands
is contaminated by the emission from various spectral lines.  The
strongest spectral line emission that we detect is the one from the
CO~3--2 transition located at a rest frequency of 345.796~GHz. In
addition, there is spectral line emission from various molecules
such as SiO~(8--7 transition) and H$^{13}$CO$^+$~(4--3 transition),
as well as SO$_2$ and CH$_3$OH. There was also emission from other
molecular transitions, but these were significantly weaker.  The
CO and SiO spectral lines are good tracers of the outflow activity
in the earliest stages of star formation, whereas the \htcop\ is a
good tracer of the dense circumstellar gas. To create a pseudo-continuum
channel for each sideband the contribution from the spectral lines
was removed and the data were then averaged over all the spectral
channels. These spectral line free and polarization calibrated data
were used to produce maps of the I, Q, and U Stokes parameters.
These maps were then independently deconvolved using the CLEAN
algorithm. Since the continuum emission (Stokes I) from this source
is quite strong,  the source visibilities could be self-calibrated.
The gain solutions from the self-calibration were applied to all
the Stokes visibilities (continuum and line emission).  The Q and
U maps were combined to produce maps of the debiased linear
polarization intensity \citep{Leahy89}, the fractional polarization,
and the position angle.  Continuum maps were made in each sideband
separately for each of the observing dates and these were identical
within the limits of the noise.  Combined maps were then obtained
using with the two sidebands and the two observing dates. Stokes
I, Q and U maps of CO~3--2, SiO~8--7, and H$^{13}$CO$^+$ 4--3 were
also obtained.  For the spectral lines, we present only the Stokes
I line emission, since no significant polarized line emission was
detected.  Table~\ref{tobserva} lists the basic parameters of the
resulting maps, including the frequency of the spectral lines or
continuum, the channel resolution (for line observations), the
resulting synthesized beam, and the root mean square ($rms$) noise
of the maps.

\section{Results}

\subsection{Continuum Emission}

\begin{figure}
\includegraphics[scale=0.7]{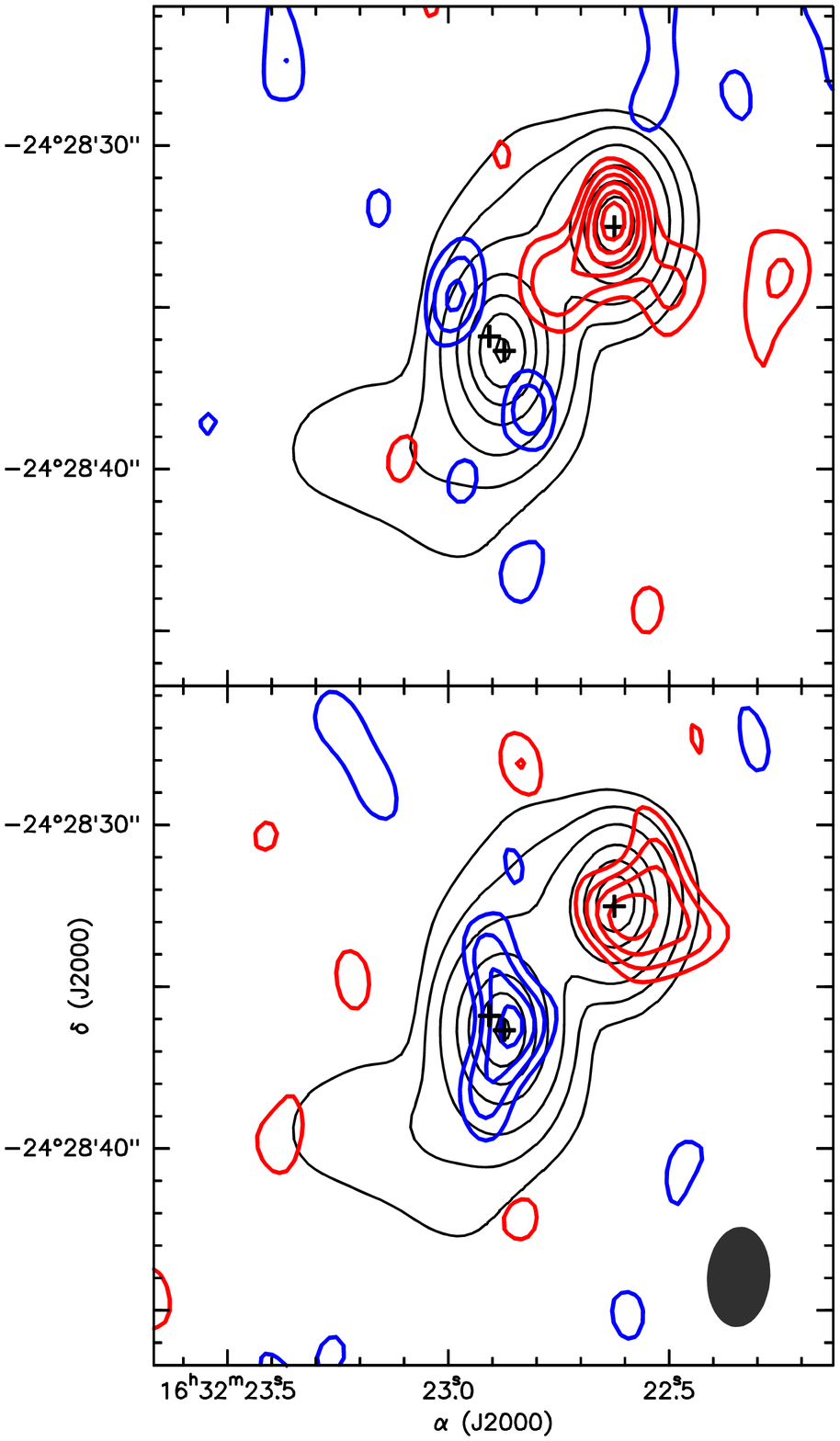}
\caption{ {\em Top panel:} Color contour map of the Stokes U dust
emission overlaid on the black contour map of the total dust emission
(Stokes I). Stokes U blue (negative) and red (positive) contours
are $-5$, $-4$, $-3$, $-2$, 2, 3, 4, 5, 6, and 7 times the $rms$
noise of the map, 4~mJy~beam$^{-1}$. Stokes I contours are 2, 7,
17, 37, 57, 77, and 97\% of the peak intensity (4.1~Jy~beam$^{-1}$).
Crosses mark the position of Aa, Ab, and B sources \citep{Loinard07}.
The synthesized beam is shown in the bottom right.  {\em Bottom
panel:} Same as previous panels, but with Stokes Q in blue (negative)
and red (positive) contours.
\label{IQU}}
\end{figure}

The Stokes I map of the total continuum flux density is shown in
Figure~\ref{IQU}.  The continuum emission, which arises from the
warm dust, is quite strong and is resolved into the two well known
components, sources A \& B \citep{Wootten89}. Subarcsecond resolution
observations at $\lambda \sim 1$ mm show that  source A itself is
comprised of two components, Aa and Ab, separated by 0.6$\arcsec$
\citep{Chandler05} with Aa located southwest of Ab. In addition,
subarcsecond VLA observations show that the centimeter (cm) wavelength
emission from Aa can be further split into two additional components
\citep{Loinard07}.  The peak of the Stokes I continuum is $\sim4.1$
and $\sim3.9$~Jy~beam$^{-1}$ for source A and B respectively, and
the integrated flux from both sources is approximately 11.5~Jy.
The absolute flux scale at the SMA is only determined to an accuracy
of 5\% and thus all the flux densities calculated by us are uncertain
by this factor. Single dish observations with the JCMT at 850~$\mu$m
($14\farcs6$ beam) measure a flux of $23.6\pm1.3$~Jy within a radius
of $20\arcsec$ of the two sources \citep{Correia04}.  The SMA
measurements appear to detect about half of the total flux.  This
is due to the fact that the SMA antennas have a primary beam size
(diameter) of 34$\arcsec$ at the chosen observing frequency and the
sensitivity to extended structure degrades as the distance from the
pointing center increases.  This missing undetected flux must
therefore arise from emission on size scales larger than $\sim15\arcsec$
which is about six times the synthesized beam size.

A Gaussian fit to source B shows that is barely resolved (see
Table~\ref{tbl-props}), with a radius less than 100~AU.  This is
consistent with previous higher angular resolution observations
which indicate that the dust emission comes from a optically thick
disk with an outer radius of 26~AU  \citep{Rodriguez05, Loinard07}.
The measured flux of source B is in agreement with the value measured
at at frequency of 305~GHz by \citet{Chandler05} with higher angular
resolution, after taking into account its spectral index. The
Gaussian fit to source A shows that this source is more extended,
and is resolved with a deconvolved scale of $330\times180$~AU,
elongated in the north-south direction (see Table~\ref{tbl-props}).
The flux measured is $\sim 1$~Jy higher than that expected from
\citet{Chandler05}, probably because they filter some emission in
their higher angular resolution maps. In addition to the contribution
from sources A and B, there is some contribution from more extended
dust, mostly from the southeastern part of source A and also north
of source A (east of source B). The total flux measured with the
two Gaussian fits of is $\simeq 10.6$~Jy (Table~\ref{tbl-props}),
whereas the total flux measured with the SMA is 11.5~Jy.

\subsection{Dust Polarization \label{ResultsDustPol}}

\begin{figure}
\includegraphics[scale=0.7]{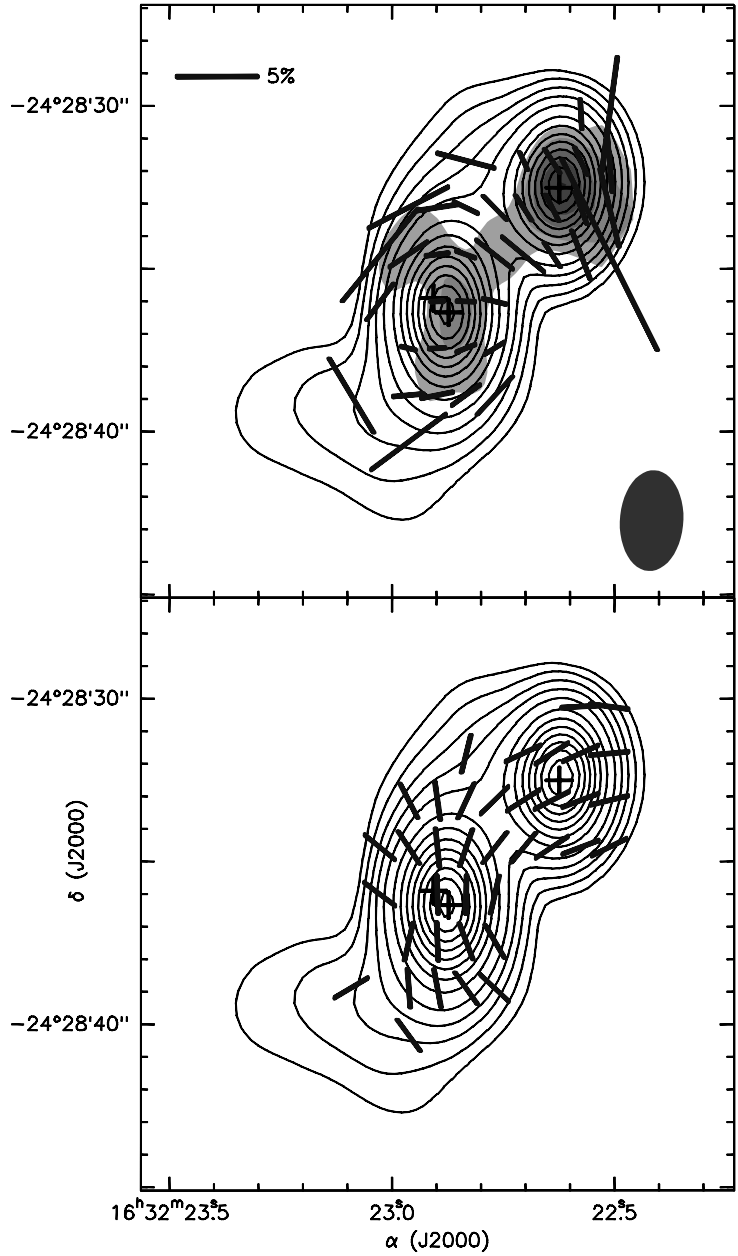}
\caption{ {\em Top panel:} Contour map of the total (Stokes I) dust
emission overlaid on the grayscale image of the polarized dust
intensity.  The gray bars represent the polarization vectors.  The
5\% vector length is shown in the top left panel for comparison.
Their length is proportional to the polarization fraction. Contours
are  2, 4, 7, 11, 18, 28, 38, 48, 58, 68, 78, 88, and 98 percent
of the peak intensity (4.1~Jy~beam$^{-1}$). Black crosses mark the
position of Aa, Ab, and B sources \citep{Loinard07}. The synthesized
beam is shown in the bottom right.  {\em Bottom panel:} Contour map
of the total dust emission as in the top panel. The bars represent
the magnetic field vectors.
\label{Bmap}}
\end{figure}

The linearly polarized component of the emission can be obtained
from maps of Stokes Q and U. Typically, this is quite small and is
only a few percent of the Stokes I emission.  The maps for Stokes
Q and U are plotted in in the top and bottom panels of Figure~\ref{IQU}.
The peak (absolute) values of Stokes Q and U  are $\sim$ 7 times
the noise level of $\sim$ 4~mJy~beam$^{-1}$. Note that in contrast
to Stokes I which is a positive quantity, Q and U can be negative.
We then obtained the maps of the (debiased) linearly polarized flux
density ($P$), the polarization position angle ($\theta$), and the
fractional polarization ($p$) which is expressed as a percentage.
The maps of the errors in $P$, $p$, and $\theta$ are obtained as
well.  The map of the polarized intensity, fractional polarization,
and position angle overlaid on a map of the total intensity is shown
in the top panel of Figure~\ref{Bmap}. The fractional polarization
and position angle are only computed at points where the debiased
polarized flux density is greater than 8~mJy~beam$^{-1}$ ($\sim
2\sigma$).  Table~\ref{tbl-listpol} contains a listing of the
polarizations measured at various locations on the map.
The errors in
fractional polarization and position angle depend inversely on the
Stokes I flux density and the polarized flux density respectively.
Consequently, the errors in the fractional polarizations are smaller
in regions where the continuum flux density is higher, while the
position angle errors are smaller where the polarized flux density
is larger.  

From the map of the polarized emission (top panel of Figure~\ref{Bmap})
we can see that the polarization structures and morphologies are
considerably different for the two sources A and B.  
The position angles of the linear polarization around
source A appear to be approximately in a ``centrosymmetric'' pattern.
Such a  pattern can also be produced when the polarization is caused
due to scattering \citep{Silber00}. The scattering cross-section
of the dust grains is quite significant in the optical and near-IR
bands. However, it is inversely proportional to $\lambda^4$ and
thus the scattered polarized radiation decreases rapidly as the
wavelength increases.  Therefore, the contribution from scattering
is likely to be negligible at the much longer submm observations.
The polarization that is observed by us with the SMA must therefore
arise largely from the continuum dust emission.

\begin{figure}
\includegraphics[scale=0.9]{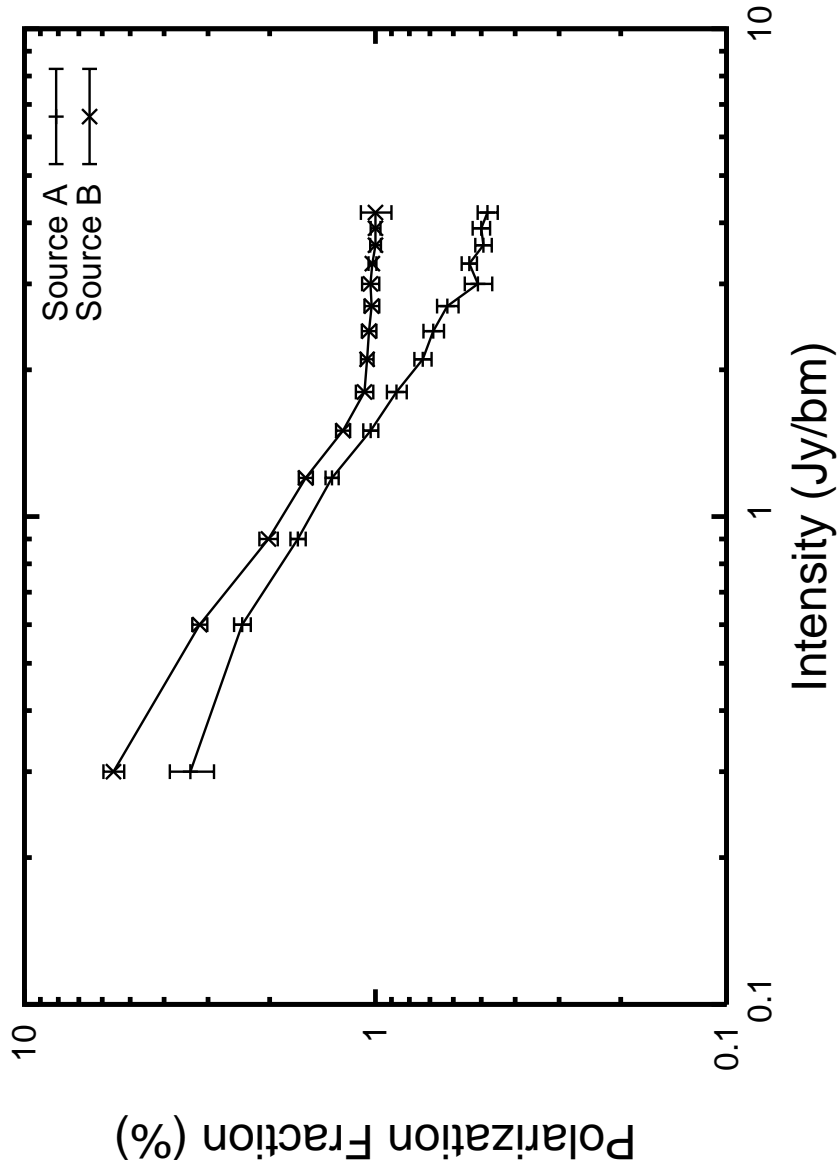}
\caption{The fractional polarization as a function of total intensity
for each of the two sources A and B. The decrease in fractional
polarization at large continuum Stokes I (also called a ``polarization
hole'') has also been seen at other polarization observations as
well \citep{Schleuning98}. For both sources the errorbars are also
plotted and the fractional polarization errors are quite small
compared to the value.
\label{p-vs-I}}
\end{figure}

The polarization fraction for the emission from source B is higher
than that of source A, as seen in Figure~\ref{p-vs-I}.  Furthermore,
this plot shows that the decrease in fractional polarization at
larger values of the total intensity is greater for source A.  This
decrease in fractional polarization towards the center of the cores
where the peak emission is higher indicates that there may be a
``polarization hole'' effect.  This decrease or depolarization is
also seen in other sources as well \citep{Schleuning98,Matthews01,Lai02}.
This effect could be due to at least three possible factors. Firstly,
the higher density and temperature towards the center of the sources
can lead to a misalignment due to the effects of a higher collisional
rate and hence lower the degree of polarization.  Secondly, the
grain properties towards the center could be different from the
grain properties in the outer envelope with the grains in the center
being less able to align with the field than the grains in the
envelope.  The difference in the decreases in sources A and B could
possibly be due to differences in the grain growth and properties
in the two sources.  One other possibility, which our observations
seem to hint at, is that this decrease is due to the significant
disparity in the morphology of the magnetic field. This central
depolarization was also suggested by \citet{Matthews01} from their
observations of the OMC-3 filaments in Orion A.  Towards the center
of source A, the magnetic field directions change significantly
over small scales leading to spatial variability in the position
angles of the polarized emission.  The resolution of our observations
is not  sufficient to resolve the small scale structure.  This
results in lower polarization at the center due to the averaging
of small scale structure.

The polarized emission from this source has been the target of a
number of different observations conducted both with single dish
telescopes and interferometer arrays at a range of wavelengths
(Table~\ref{compols}). As mentioned earlier in the introduction,
most of the early measurements  were inconsistent with each other,
which was perhaps due to limited sensitivity.  A more recent
polarization map from the JCMT using the Submillimeter Common Use
Bolometer Array (SCUBA) polarimeter shows that at the peak of the
dust continuum emission, no polarization is detected \citep{Matthews09}.
We can compare our observations by convolving our images by the
beam equal to the resolution of the JCMT. Since almost all of the
area over which the polarized dust emission detected by the SMA is
within the JCMT beam, this is approximately equal to the integrated
polarization over our map. When this is done we obtain extremely
low polarizations around ~0.2\% (Table~\ref{compols}), and  is in
good agreement with the SCUBA measurements.

Under the assumption that the grains are not spherical and are
rotating about their short axis which is aligned with the magnetic
field, the field structure can be obtained by rotating the position
angle of the observed polarization by $90\arcdeg$ (bottom panel of
Figure~\ref{Bmap}). There is considerable spatial structure in the
deduced magnetic field directions, with a large scale twisted
magnetic field in a direction coinciding with a curve joining sources
A and B. In addition, source A shows that the lines are deformed
with an ``hourglass'' like structure.  This is the second such
sensitive detection of this type of ``hourglass'' structure towards
a region of low mass star formation. The first was the young stellar
object IRAS~4A reported by \citet{Girart06}. In contrast, source B
shows very little variation in magnetic field structure.

\subsection{Molecular Lines}

\subsubsection{\htcop\  4--3}

\begin{figure}
\includegraphics[scale=1.0]{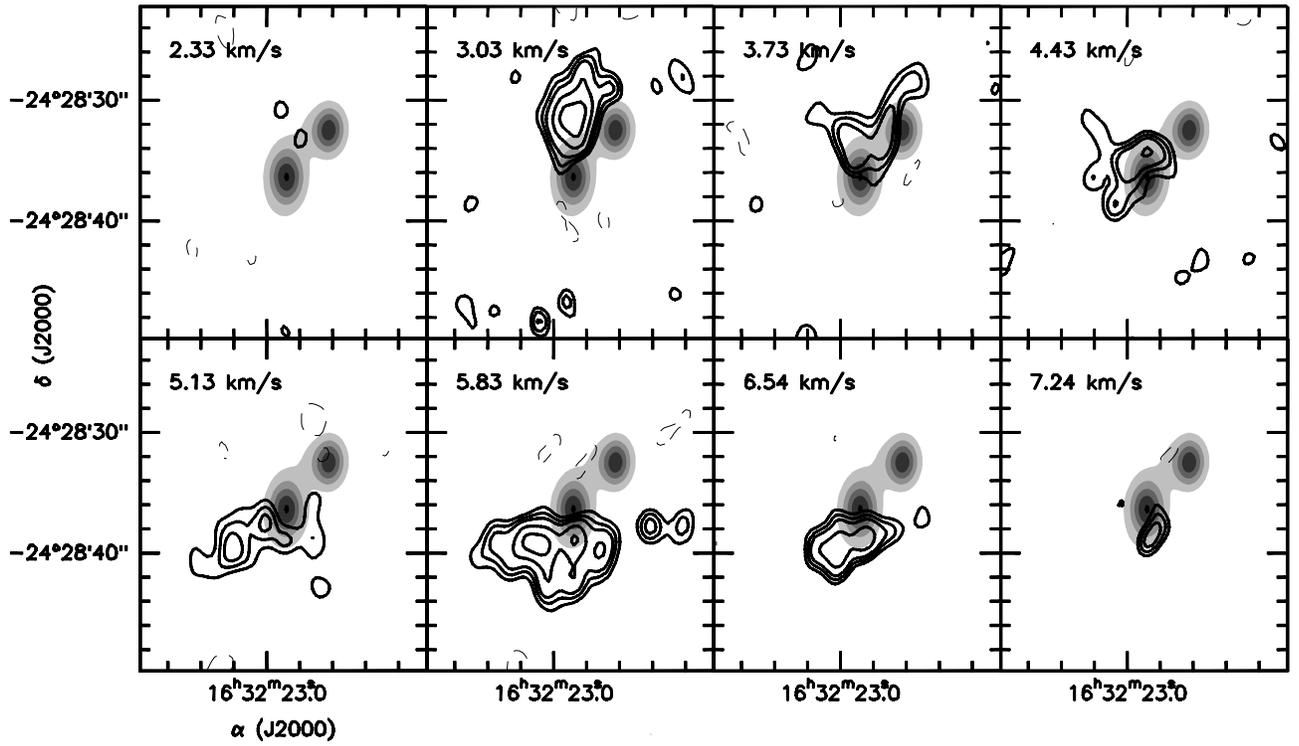}
\caption{
Contour channel maps of the \htcop\ 4--3 line emission overlaid on
the dust emission which is shown in the gray scale.  The contour
levels are $-3$, 3, 4, 5, 7 and 9 times the $rms$ noise level,
0.42~Jy~beam$^{-1}$. The $v_{\rm LSR}$ velocity of each channel is
shown in the top left corner of the panels.
\label{h13copA}}
\end{figure}

\begin{figure}
\includegraphics[scale=0.6]{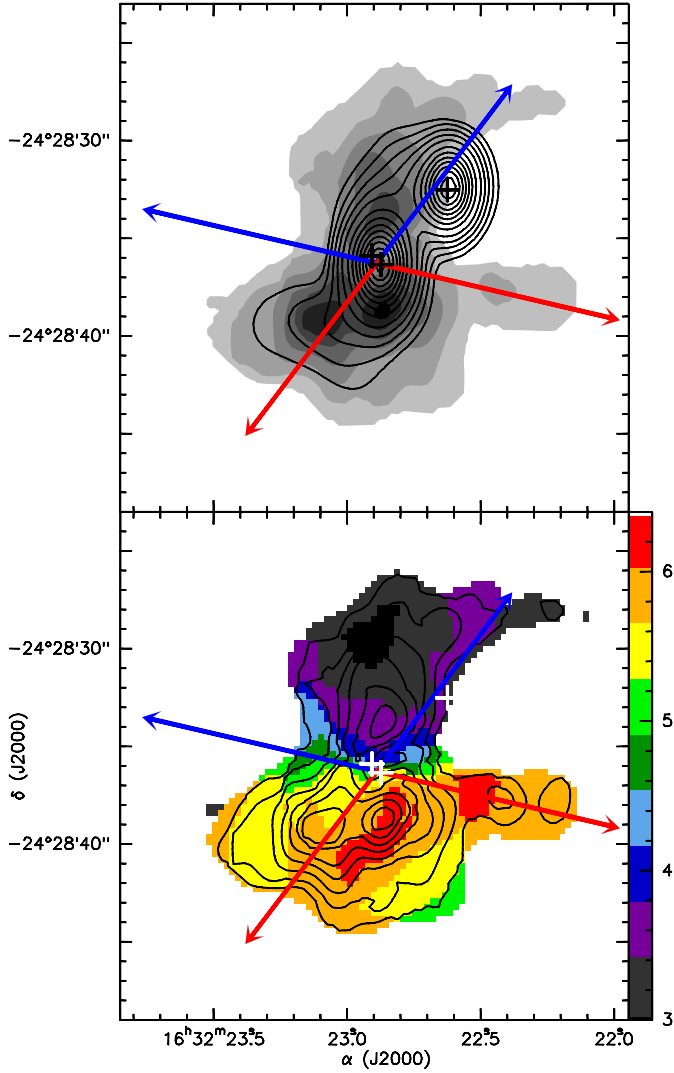}
\caption{ {\em Top panel:} Contour maps of the dust emission overlaid
with the integrated intensity (zeroth moment) gray scale image of
the \htcop\ 4--3 line emission.  Contours levels are the same as
those in Fig.~\ref{Bmap}.  Crosses mark the position of Aa, Ab, and
B. Blue and red arrows show the directions of the  E-W CO outflow
and the NW-SE SiO outflow.  {\em Bottom panel:} Contour maps of the
integrated intensity (zeroth moment) overlaid with the intensity
weighted velocity (first moment) color image of the \htcop\ 4--3
line emission. Contour levels go from 5 to 95\% the maximum value
(8.7 Jy~\kms) with steps of 15\%.  The units of the vertical bar
is in \kms.  Crosses mark the position of Aa, Ab, and B.
\label{h13copB}}
\end{figure}

Figure~\ref{h13copA} shows the channel maps of the \htcop\ 4--3
emission while  Figure~\ref{h13copB} shows the integrated emission
and the velocity field. Both figures show that the \htcop\ emission
arises roughly extended in the north-south direction over approximately
$17\arcsec \times 10\arcsec$ ($2500\times1500$ AU) and centered on
source A. The strongest emission is offset by a few arcseconds to
the north and south of source A. Indeed, the \htcop\ integrated
emission presents a relative minimum at the position of source A.
Source B appears to be devoid of the \htcop\ emission. As shown in
Figures~\ref{h13copA} and \ref{h13copB}, there are clear signs of
a velocity gradient along the major axis of the \htcop\ structure
(north-south direction) of about $\simeq0.31$~\kms~arcsec$^{-1}$,
which translates to a physical scale of 430~\kms~pc$^{-1}$, or an
angular velocity, $4.5\times10^{-12}$~s$^{-1}$. This is almost one
order of magnitude higher that the value found by \citet{Narayanan98}
from single-dish observations of the IRAS~16293 core. It is possible
that this discrepancy may be due to the different scales that are
being probed. The  \citet{Narayanan98} scale-sizes are $\sim$39$\arcsec$
which is more than twice the maximum structures being mapped by us.
The north-south velocity gradient is also observed in the
H$_2$CO~$4_{1,3}$--$3_{1,2}$ line, although this line has a slightly
different morphology: its emission arises from the two sources, A
and B, and it is extended south of source A  \citep{Chandler05}.

\subsubsection{SiO~ 8--7}

\begin{figure}
\includegraphics[scale=0.8]{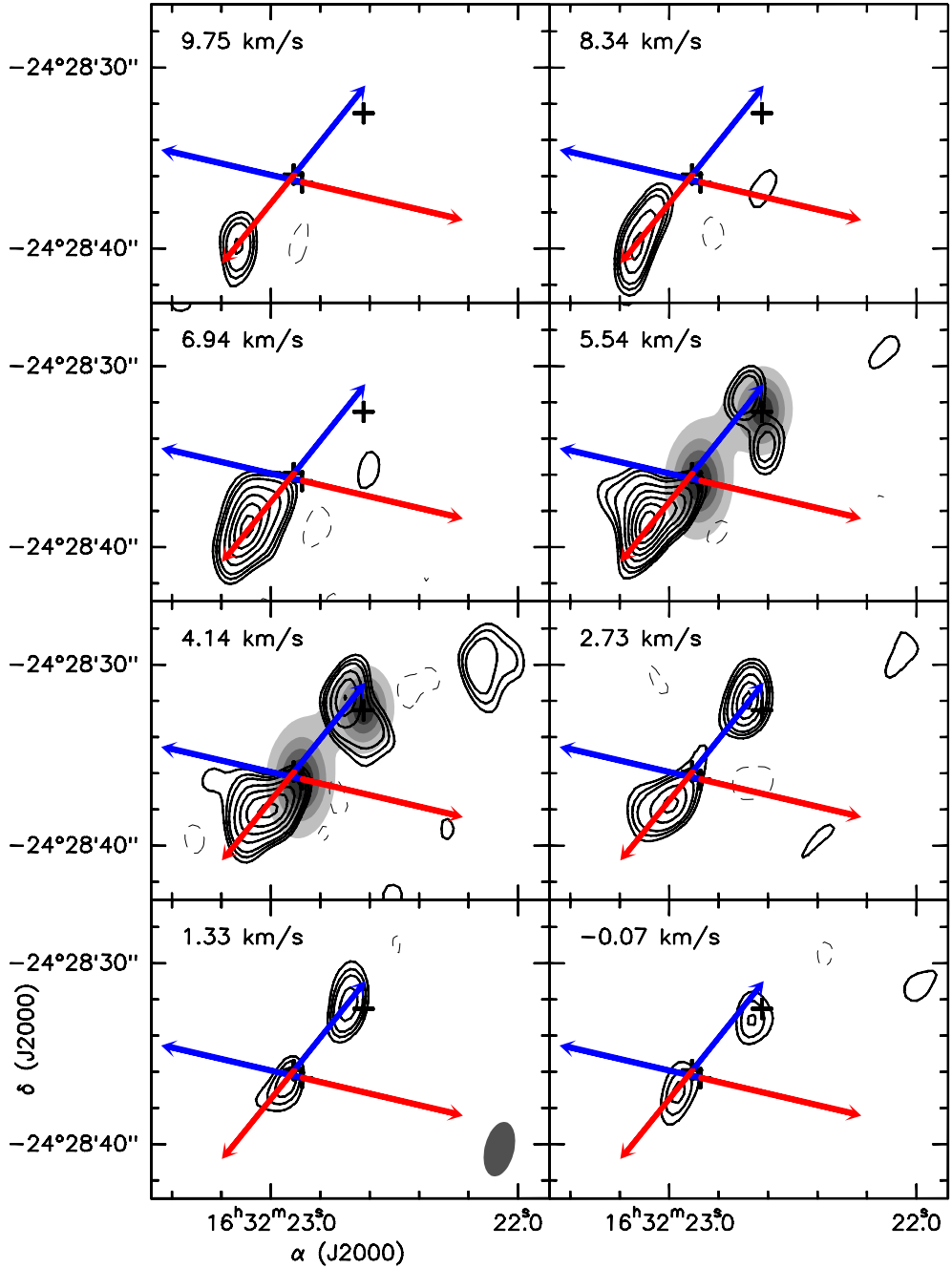}
\caption{ Channel contour maps of the  SiO~8-7 line emission. For
two of the channels, the dust continuum image in gray scale is
overlaid. Contours are  $-3$, 3, 4, 5, 7, 9, 11, 13 and 15 times
the $rms$ noise level, 0.35~Jy~beam$^{-1}$.  Crosses mark the
position of Aa, Ab and B.  The directions of the outflow is shown
as red and blue arrows. The SE-NW arrows have a position angle of
$145\arcdeg$ and are centered on source Ab, and matches well with
most of the SiO emission. The NE-SW arrows have a position angle
of $105\arcdeg$ and are centered on source Aa.
\label{sio}}
\end{figure}

Figure~\ref{sio} shows the channel maps of the SiO~8--7 emission,
which extends in velocity over 10~\kms\ and is split into three
main condensations. The most prominent one is observed just  southeast
of source A  (SE condensation), and is detected in all the displayed
channels, although it is brighter at the redshifted velocities.  A
Gaussian fit to the emission from this condensation in  the $v_{\rm
LSR} = 6.94$~\kms\ velocity channel gives a position angle of
$145\arcdeg$, which is similar to the overall orientation of this
condensation with respect to source A (see Figure~\ref{sio}). The
second condensation appears to arise northwest of source A (and
southeast of source B) at systemic and blueshifted velocities (NW--1
condensation). At the $v_{\rm LSR}$ of 5.5~\kms\ the emission appears
to break up  into two condensations (also partially observed at the
4.14~\kms\ velocity channel). The peak intensity of the brightest
channels appears to be located in the same axis as the one formed
between SE condensation and source A. Indeed, the line that connects
the SE and NW condensation passes closer to Ab than Aa. The third
condensation is located about 6$\arcsec$ northwest of source B
(NW--2 condensation).

\subsubsection{CO~ 3--2}

\begin{figure}
\includegraphics[scale=0.9]{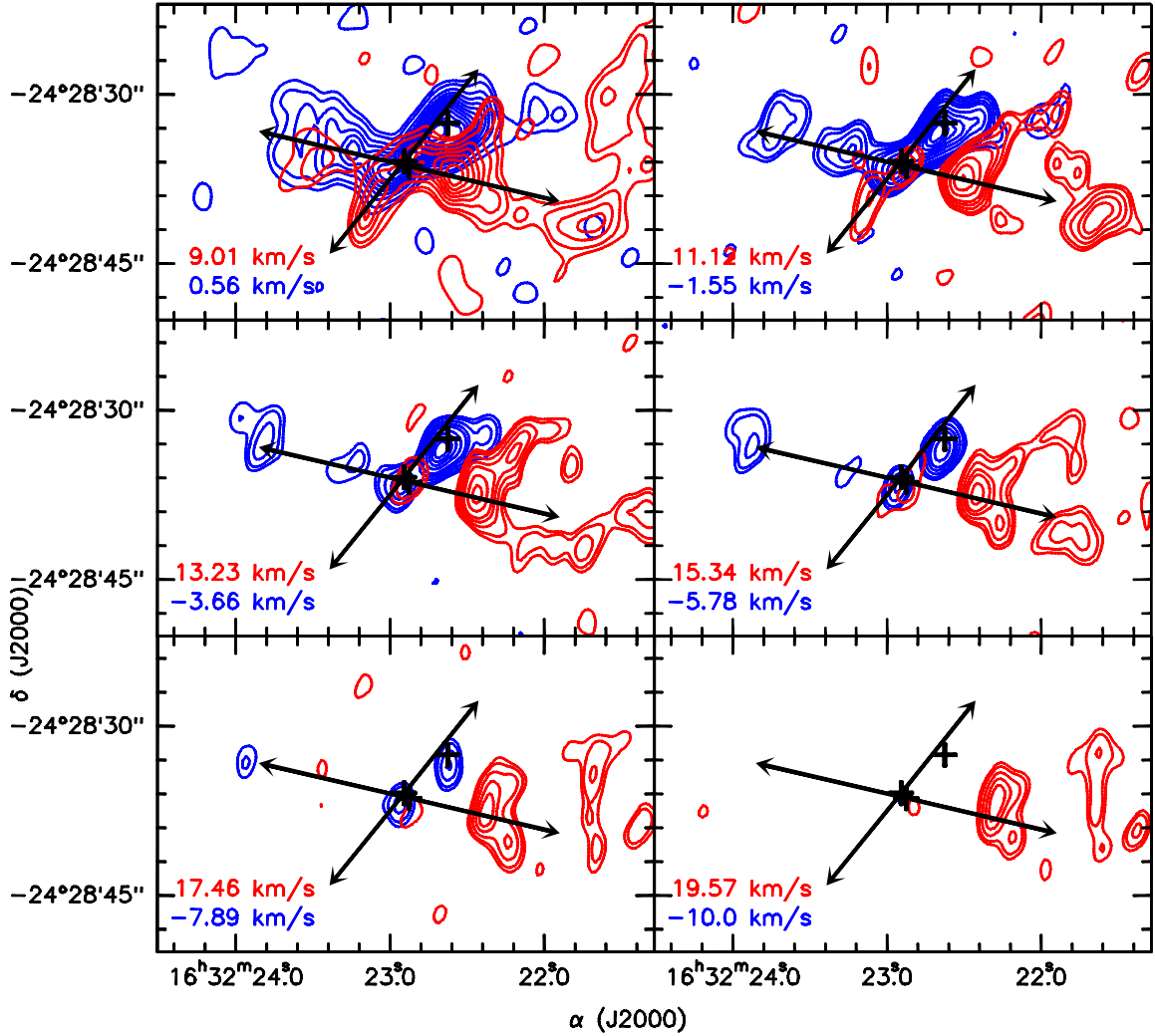}
\caption{Channel maps of the CO~3-2 for the blueshifted (blue
contours) and redshifted (red contours) emission. For the top-left
panel the contours are 5, 10, 20, 30, $\dots$ 120 times the $rms$
noise level, 0.25~Jy~beam$^{-1}$. For the other panels, the contours
are 4, 6, 10, 15, 20, 30, 40, 50, 60, 70 and 80 times the $rms$
noise level.  The $v_{\rm LSR}$ velocity of each channel is indicated
in the top left part of each panel.
\label{Outflow1}}
\end{figure}

The high angular resolution maps of the CO emission (for the 2--1
and 3--2 lines) from SMA observations have been already reported
in the literature \citep{Yeh08}. The emission is quite extended in
the east-west direction (E-W outflow). In the visibility domain of
our CO~3--2 data set, this shows up as a steep increase of the CO
flux for visibilities with a radius, $r_{u,v}$, shorter than
20~k$\lambda$. Figure~\ref{Outflow1} shows the channel maps of the
CO~3--2 obtained by excluding the visibilities with $r_{u,v} <
20$~k$\lambda$. The most prominent emission comes from the E-W
outflow. The blueshifted eastern lobe is more collimated than the
redshifted western lobe, which has an open shell structure.  The
position of peak intensity of the brightest clumps (both eastern
and western) appear to be well aligned, crossing source A, with a
PA$\simeq105\arcdeg$.  The SiO outflow is also traced by the CO
3--2 emission. The SiO SE condensation is detected in the lowest
redshifted velocity channels ($v_{LSR}=9.0$ and 11.1~\kms). There
is also blueshifted CO emission apparently associated with the SiO
NW~1 condensation, although the emission appears to be slightly
displaced to the west, with the emission even being more bent to
the west of source B.

\section{Analysis}

\subsection{Distribution of Polarization Position Angles}

\begin{figure}
\includegraphics[scale=0.7]{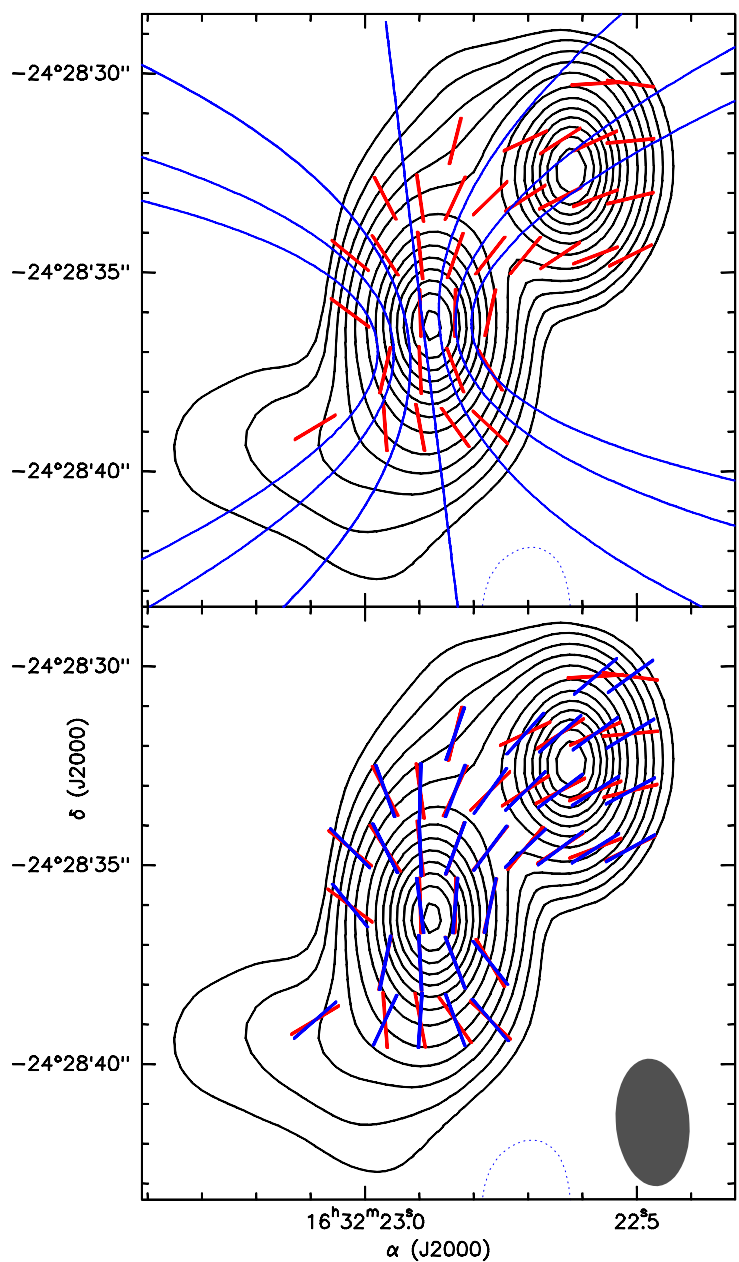}
\caption{ {\it Top panel:} Overlap of the 0.88~mm dust emission and
the B-vectors (same as Figure~\ref{Bmap}) with the best solution
for the parabolic family of functions (blue solid lines).  {\it
Bottom panel:} Same as top panel, but the best family of functions
are shown as B-vectors (in blue) at the same position as the measured
values.
\label{Bparabola}}
\end{figure}

The average values of the polarization PA around source A and B are
$95\arcdeg$ and $24\arcdeg$ respectively.  The distribution of the
residual values of the polarization PA after subtraction of the
average values around source A and B are shown in the bottom panel
and the middle panel of Figure~\ref{PAdist} respectively.  This
distribution shows that source B has a more uniform pattern than
source A, which is likely due to the deformed morphology in the
east-west direction (see Figure~\ref{Bmap}). In spite of the magnetic
field structure, the analysis of the dispersion of the position
angle as a function of displacement through the ``structure function''
can provide an indirect measurement of the turbulent to magnetic
energy ratio \citep{Hildebrand09}. Since the range of scales where
the polarization is detected is small, we prefer to fit the magnetic
field morphology around source A in a similar way as was done in
IRAS~4A \citep{Girart06}.  As a first approximation of the field
geometry we fit the magnetic field vectors around source A only,
excluding those around source B, with a set of parabolas. The results
obtained with this method are in fair agreement with a more detailed
analysis using specific theoretical magnetic field geometries
\citep{Goncalves08}.  The parabolic functions used are of the type:
\begin{equation} (y-y_0) = g_i + g_i \, C (x-x_0)^2 \end{equation}
where $y_0$ and $x_0$ is the center of symmetry of the magnetic
field configuration. The quantity $g_i$ is a parameter which depends
on the curve that is selected and $g_i C$ represents the quadratic
term. Note that the main magnetic field direction is along the $x$
axis, and that  for the observed magnetic field morphology, the $y$
axis is going to be close to the right ascension axis. We used the
$\chi ^2$ method with $y_0$, $x_0$, $C$ and the position angle of
the main direction of the magnetic field, $\theta_{\rm B}$ as free
parameters. The best fit solution obtained is with $x_0=-0\farcs10$,
$y_0=-0\farcs62$, $C=0.25$ and  $\theta_{\rm B} = 7.0\arcdeg$.
Figure~\ref{Bparabola} shows the best solution for five sets of
parabolas (top panel), and the values of the modeled magnetic vectors
at the position of the observed magnetic field vectors (bottom
panel). It is remarkable that this solution not only fits the
magnetic field vectors around source A, but those around source B
as well even though these were not used in the fitting!  The
distribution of the residuals, including the magnetic field vectors
from source B, is shown in the top panel of Figure~\ref{PAdist}.
The standard deviation of the residuals is $\delta \theta_{obs}=13.9\arcdeg
\pm 0.6\arcdeg$. The uncertainty of the polarization position angle
is $\sigma_{\theta} = 9.8\arcdeg \pm 3.8\arcdeg$, so the intrinsic
dispersion is $\delta \theta_{\rm int} = 9.8\arcdeg \pm 3.9\arcdeg$.

\subsection{Physical Parameters: Mass, Density and Column Density}

We can infer the total mass from the intensity of the dust emission
using the following relation (Hildebrand 1983) 
\begin{equation} M
= \frac{F d^2}{\kappa B(T)} 
\end{equation} 
where $F$ is the flux, $d$ is the distance to the source, $\kappa$
is the dust opacity, and $B(T)$ is the Planck function.   By fitting
models to the single dish data obtained in the far infrared and
submm regime, \citet{Correia04} estimate the dust opacity law
coefficient to be  $\beta \simeq 1.8$. From their results, we
estimate the optical depth at 880~$\mu$m to be less than 0.18 at
size scales greater than 150 AU (which is 95\% of their fitted cloud
size of 3000 AU).  Thus, the emission is fairly optically thin over
most of the cloud. This, however, is not the case towards the center
of source B \citep{Loinard08}.  The dust temperature at the observed 
scales,  is taken to be T$ \sim 50$~K  \citep{Correia04}.

The integrated flux from sources A and B  are given in
Table~\ref{tbl-props}. The calculated masses of sources A and B are
0.33 M$_\odot$ and 0.22 M$_\odot$ respectively and these values are
similar to ones derived by \citet{Chandler05}. However, for source
B, this mass is a factor of 2 lower than the value found by
\citet{Rodriguez05}, and could possibly be due to the fact that the
dust emission comes mainly from an optically thick disk. In order
to determine the column density and the number density, we also
need the total area used to estimate the flux density. From this
area, we can define an equivalent radius of $R = \sqrt{{\mathrm
Area}/\pi}$. Following the same approach taken in the case of IRAS~4A
\citep{Girart06}, we select the  area of emission to be the entire
region over which we can detect the continuum flux density (Stokes
I).  The equivalent radii, as calculated from the areas, are
4$\arcsec$ and 3$\arcsec$  for sources A and B respectively. The
column and volume densities source A and source B are given in
Table~\ref{tbl-props} and is of the order 10$^{23}~\mathrm{cm}^{-2}$
and 10$^7~\mathrm{cm}^{-3}$ respectively.  The combined total mass
of this system according to our measurements is 0.55 M$_\odot$,
with a column density of 6.4$\times10^{23}~\mathrm{cm}^{-2}$ and a
volume density of 4.7$\times10^7~\mathrm{cm}^{-3}$. As discussed
earlier, the total mass calculated by us  is lower than the values
determined from single dish measurements which are more sensitive
to the emission on much larger scales. The mass of the envelope can
be approximately determined if we assume that it contributes most
of the missing flux of approximately 12~Jy. Using a value of 20~K
for the  envelope temperature, we obtain an envelope dust mass of
approximately 1.9~M$_\sun$. This calculated mass is comparable to
those derived from other observations of this source at a number
of different wavelengths. These observations show that the envelope
dust mass is  in the range of 2--3~M$_\sun$
\citep{Walker90,Mezger92,Andre94,Correia04}, which is approximately
80\% of the total mass of the system. This ratio is not very different
from that seen in other Class~0 sources such as IRAS~4A
where its value is almost 90\% \citep{Jorgensen07}. The larger
amount of envelope mass in IRAS~4A indicates that it is likely to
be not as evolved as IRAS~16293.

From the observed \htcop\ velocity gradient, we can derive the
dynamical mass needed for equilibrium between the gravitational and
centrifugal forces: $M_{\rm dyn}  = v_{\rm rot}^2 \, R \, /G $,
where $v_{\rm rot}$ is the rotation velocity and $R$ is the radius
of the flattened structure. For the measured values, $R=5\farcs2$
(780~AU) and $v_{\rm rot}=3.31 / {\rm sin}^2 i$~\kms\ ($i$ is the
inclination angle of the rotation axis with respect to the line of
sight), the dynamical mass is $M_{\rm dyn}  =$~0.084~sin$^{-2}
i$~M$_{\sun}$. Assuming that the rotation axis and the outflow axis
are parallel (as they appear to be in projection), then the inclination
angle derived from the outflow is $i \sim 50$--$60\arcdeg$
\citep{Yeh08}, thereby $M_{\rm dyn}  \simeq$~0.11--0.14~M$_{\sun}$.
Therefore, the circumstellar mass around source A plus the mass
already accreted onto the protostar is larger than the dynamical
mass, so this flattened structure is not stable, and likely is
undergoing collapse, and has been inferred from spectral signatures
of infall \citep{Chandler05, Remijan06, Takakuwa07}.

\subsection{Physical Parameters: Magnetic Field Properties}

The magnetic field strength can be estimated indirectly using two
different methods: from the modified Chandrasekhar-Fermi (C-F)
equation \citep{Chandra53,Heitsch01}; and from the curvature of the
deformed, ``hourglass''-like, field lines around source A.  

\subsubsection{Modified Chandrasekhar-Fermi Method}

The modified C-F equation is:  $B  = Q \, (\delta v /  \delta \theta)
\,  \sqrt{\rho}$, where $\delta v$ is the velocity dispersion along
the line of sight, $\delta\theta$ is the intrinsic dispersion in
the polarization position angles, $\rho$ is the volume density, and
$Q$ is a dimensionless parameter that depends on the relative
strengths of the magnetic field and the turbulence. We adopt a value
of  $Q=0.5$, which is appropriate for turbulent magnetized clouds
with relatively strong fields, $\delta \theta \la 25\arcdeg$
\citep{Ostriker01}.  We use the values derived in the previous
section for the volume density and the intrinsic dispersion $\delta
\theta$.  The velocity dispersion is obtained from the H$^{13}$CO$^+$
4--3 spectral line emission as this spectral line approximately
traces the same spatial scale as the polarization that is detected
by us.  From the intensity weighted velocity dispersion (second
moment) map, the line of sight velocity dispersion is $\delta
v\simeq0.35$~\kms\ (i.e., a FWHM of 0.82~\kms)  in the regions where
the emission is strong and is not affected by the strong velocity
gradient seen in the north-south direction around source A. Then,
using the modified C-F expression given in \citet{Lai02}, we find
that the component of the magnetic field strength on the plane of
the sky is $\simeq4.5$~mG.

\subsubsection{Magnetic Field from Curvature of Field Lines}

The gravitational collapse of the cloud (neutral and ion particles)
pulls the field lines into the canonical ``hourglass'' shape,
producing a magnetic tension force resisting the collapse. This
force, which is proportional to ${\bf (B \cdot \Delta) B}$, can be
approximately expressed as $B^2/R$, where $R$ is the radius of
curvature.  If the gravitational force is known, it is possible to
estimate the magnetic field strength from the observed curvature
of the field lines using the following equation as derived from the
expressions given by  \citet{Schleuning98}.
\begin{equation}
\left[ \frac{B}{\rm 1 \, mG} \right]^2 = \left[ \frac{R}{\rm 0.5
\, pc} \right] \left[ \frac{D}{\rm 0.1 \, pc} \right]^{-2} \left[
\frac{M}{\rm 100 \, M_{\sun}} \right] \left[ \frac{n{\rm (H_2)}}{\rm
10^5 \, cm^{-3}} \right] 
\end{equation} 
where $D$ is the distance of a field line from the protostar. The
radius of curvature of the magnetic field lines can be estimated
from the family of fitted parabolic functions.  For a parabola, $y
= a \, x^2 + b$, the radius of curvature at the origin of the
abscissa, is $R=1/(2 \, a)$.  We selected the distance from the
protostar to be 2$\arcsec$ as this is approximately equal to our
resolution along that direction. Furthermore, at larger distances
along the center of source A, no polarization vectors are significantly
detected.  At this distance of 2$\arcsec$ (300~AU), the radius
of curvature of the field line around  source A is $\simeq 0\farcs77$
(116~AU). The volume density  measured previously is $n{\rm
(H_2)}\simeq 4.9\times 10^7$~cm$^{-3}$ and the circumstellar mass for
source A $\sim 0.33$~M$_{\sun}$.  With these numbers, the magnetic
field strength required for the observed curvature is about 3.5~mG,
which is in reasonable agreement with the value estimated from the
C-F method.

Henceforth, we will estimate the relevant physical quantities using
a field strength of $\sim$4.5~mG.  Using the estimated average
column density, we can calculate the mass-to-magnetic flux ratio
\citep{Mouschovias76} to be approximately equal to the critical
value. This mass-to-magnetic flux ratio does not take into account
the mass that is already accreted onto the protostars.  From the
modeling of the polarization pattern in IRAS~4A \citep{Goncalves08},
the accreted mass in that source was similar to the mass as calculated
from the dust. Therefore, our data suggests that the cores are
approximately supercritical, which is in agreement with the fact
that this is an active star forming site. The intrinsic dispersion
of the polarization angles yields a turbulent to magnetic energy
ratio of $0.32\pm0.27$, which suggests that the magnetic energy
dominates over the turbulent energy. Finally, we can also estimate
the relationship between the angular momentum and the magnetic
fields, which are also thought to play an important role in the
dynamics of the collapse \cite{Machida05}.  If the ratio between
the angular velocity and the magnetic flux, $\omega/B$, is larger
than a critical value which can be expressed as, $(\omega/B)_{\rm
crit} = 3.21\times10^{-8}\, c_{\rm s}^{-1}$~yr$^{-1}$~$\mu$G$^{-1}$,
then the angular momentum controls the collapse ($c_{\rm s}$ is the
sound speed in \kms). If it is smaller, then the magnetic field
dominates over the centrifugal forces. The measured magnetic field
strength and the velocity gradient found in the \htcop yields a
ratio of $\omega/B =  2.8$--$4.8\times10^{-8}$. This is slightly
smaller than the critical value which is approximately $5.9\times10^{-8}$
(for $c_S=0.54$~\kms).  While the magnetic field appears to  dominate
energetically over the turbulence, it is comparable in magnitude
to the centrifugal energy.

\section{Discussion}

\subsection{Molecular Outflows, Kinematics,  and the Evolutionary
Stages of the IRAS~16293 Sources\label{Kine}}

The IRAS~16293 region shows significant outflow activity with
multiple outflows where the largest outflow is approximately in an
east-west direction. Our high-resolution CO~3--2 observations of
this outflow are in agreement with previous observations \citep{Yeh08},
that is, it is centered in source A, with an orientation of PA$\simeq
105\arcdeg$ and its morphology in the red lobe suggests that this
outflow is not very collimated (our maps were done excluding the
short baselines thus probing extended structure). The Spitzer near-IR
images from the vibrationally excited H$_2$ also trace shock
structures within this outflow but at considerable distance from
this object \citep{Padgett08}.

Our observations of the high-angular resolution CO~3--2 do not show
any obvious traces of the extended NW-SE outflow, which was also
not seen by \citet{Yeh08}. Therefore, it is possible that this is
either a fossil outflow or that its powering source is presently
in a very quiescent phase (with no strong accretion taking place).
A third outflow has also been found in this region, as traced by
the SiO~8--7 emission, centered in source A and with an orientation
of PA$\simeq 145\arcdeg$.  This outflow is quite compact (less than
$5\arcsec$ or 750~AU long) and very bright in the SiO line compared
with its CO emission. The SiO~8--7 emission in the outflow is
expected to arise from dense ($\sim 10^7$~cm$^{-3}$) and hot (400~K)
molecular gas \citep{Hirano06}.  All of this indicates that the SiO
may trace a very young outflow, much younger than the two previously
reported outflows. The powering source appears to certainly be
within source A. The geometry of the outflow marginally suggests
that source Ab may the powering source of this young outflow.  Higher
angular resolution observations are needed to attest the origin of
this outflow. The structure of the NW blueshifted lobe at systemic
velocities, which is split into two condensations with source B
roughly in between, suggests that this outflow may be partially
interacting with source B.

The emission from this new SiO outflow can also be seen in the
highest velocity channels of the combined SMA and JCMT maps of
\citet{Takakuwa07}.  They mention that the emission that they detect
at high HCN 4--3 velocities is in agreement with the CO maps. The
CO emission is quite complex because of contributions from both the
EW and the new SE-NW outflow.  The HCN 4--3 emission at ambient
velocities is possibly contaminated by this NW-SE outflow. Furthermore,
the  HCN 4--3 emission is also affected by the deep self-absorption
and the high optical depths. Thus, it may not be a good tracer to
try study the kinematics of the circumstellar/binary environment.
In contrast, the emission from the SiO spectral line detected by
us is much less contaminated and therefore, is a better tracer of
this new outflow.

Source B does not appear to show any active outflow and  the only
hint of any outflow activity may be the presence of a free-free
component at very small scales ($\sim 15$~AU) derived indirectly
from the spectral index map at cm wavelengths \citep{Loinard07}.
This has been interpreted to imply that source B has not yet started
the phase of significant mass loss \citep{Chandler05}.  Most of the
emission comes from a very compact, optically thick disk
\citep{Rodriguez05}, with a significantly smaller amount of material
in the envelope around it when compared with other Class~0 sources
(see \S~\ref{Compara}).  This suggests that most  of the  mass has
already been accreted onto the disk and protostar.  This raises the
possibility that source B may not be a true Class~0 protostar.  The
somewhat higher chemical richness of source A with respect to source
B, may be due to the higher outflow activity of source A
\citep{Chandler05}.  The lack of  active accretion onto the disk
from the circumstellar environment may be the cause of the non-existent
outflow activity.  It is possible that source B is (or was) the
exciting source of the extended NW-SE outflow.  Thus, source B may
be a transition object between Class~0 and Class~I, or possibly
even a Class~I object \citep[see][for definitions and descriptions
of the various classes of protostars]{Lada87,Wilking89,Andre93}.
In this scenario, the uniform pattern of the magnetic field could
be tracing the residual circumstellar envelope of source B. The
narrower line widths associated with source B as compared to A have
been interpreted as B being younger \citep{Wootten89}.  If most of
the material in source B comes from a massive disk in the plane of
the sky  \citep{Rodriguez05,Chandler05,Loinard07}, then kinematical
motion in the disk will not broad the line width.

Source A shows the more typical features of a Class~0 protostar,
with active and energetic molecular outflows surrounding a contracting
core (as inferred from the ``hourglass'' morphology).  This suggests
that it is in an active accretion phase. The \htcop 4--3 line  traces
dense molecular gas (its critical density is $\simeq 10^7$~cm$^{-1}$)
and somewhat flattened structure in the north-south direction around
source A.  This flattened envelope is rotating with its axis in the
east-west direction, which is in projection nearly parallel to the
more active and larger E-W.  Interestingly, this rotational signature
was first found by \citet{Mundy90}, who studied the kinematics in
this region using the C$^{18}$O line.  Their C$^{18}$O emission
peaks roughly at source A and is extended with a PA=150$\arcdeg$,
and with a size that is not too different from the value determined
from our \htcop\ maps. Furthermore, they obtain the C$^{18}$O
velocity gradient to be 2.4 \kms, whereas  the value calculated
from our \htcop observations is $\sim$3.0 \kms .  This agreement
suggests that both the  C$^{18}$O and \htcop spectral line emission
seem to trace the same kinematical signatures of rotation!

It is clear that the sources A and B are in different evolutionary
stages and are likely to have different ages as well. Using
submillimeter observations of the source VLA~1623, which is also
located in the Ophiuchus cloud, \citet{Andre93} find, based on
calculations of the infall rate, that VLA~1623 is a very young
Class~0 source about 6000 years old.  In contrast, recent surveys
of star formation activity find that the Class~0 sources in Ophiuchus
are at least three times older, with an estimated age of approximately
0.023 Myr \citep{Evans09}.  Even after accounting for this discrepancy,
these Class~0 protostars in Ophiuchus are extremely young when
compared to similar sources in Perseus which are 0.32 Myr old.
\citet{Evans09} suggest that either the Class~0 sources evolve
rapidly into the Class~I stage in Ophiuchus or that the transition
from Class~0 to Class~I is not continuous.

It is striking that the projected magnetic field configuration in
the plane of sky is parallel to the elongation of the flattened
structure and, thus, perpendicular to the rotation axis and the
main outflow axis.  One explanation for the lack of correlation
between the outflow directions and the magnetic field could be due
to the fact that the magnetic fields that we detect are mostly in
the envelope while the disk, from which the outflow originates,
could be decoupled from the envelope and  maybe precessing
\citep{Chandler05}.  A similar (but smaller) misalignment between
the magnetic field direction and the outflow direction was also
observed in IRAS~4A as well.  Furthermore, single-dish observations
of VLA~1623 also find that the magnetic field is perpendicular to
the outflow axis at scales of few thousand AU \citep{Holland96}.
Most theoretical models that follow the collapse of magnetized and
rotating cores assume that the magnetic field and the rotation axis
are aligned \citep{Banerjee06, Galli06}, which, obviously, is not
the case in IRAS~16293.  Some recent simulations have dealt with
the situation of an oblique magnetic field with respect to the
rotation axis \citep{Matsumoto04, Machida06}.  According  to the
simulations of  \citep{Matsumoto04}, the angular momentum component
perpendicular to the magnetic field axis is removed more rapidly
than the parallel one, so the net effect is that the rotation axis
becomes aligned with the magnetic field axis. This, however, does
not appear to be the situation in our measurements.  \citet{Matsumoto06}
have used the results from \citet{Matsumoto04} simulations in order
to predict the polarization pattern of an initially oblique magnetic
field (by $45\arcdeg$) in two scenarios, a strong magnetic field
and a weak one. As stated before, magnetic braking at the scales
relevant for the launch of the outflow (few tens of AU), induces
an alignment of magnetic field and the rotation axis.  Therefore,
the outflow, which is already parallel to the field lines, maybe
independent of the original field strength as well. On the other
hand, the situation at larger scales ($\sim 1000$~AU) is quite
different. In the case where the magnetic energy density is strong
compared to the centrifugal energy, the angular momentum is already
aligned at these scales, thereby one should expect that the
polarization observations shows the alignment of the B vectors with
the outflow.  If the centrifugal energy density is comparable to
the magnetic energy density, the situation is significantly different,
and the polarization observations can show a projected magnetic
field direction considerably misaligned with respect to the outflow.
The extreme case occurs for a specific configuration (inclination
angle with respect to the plane of sky direction), where the two
axes can be  almost perpendicular.  It is noticeable that in this
case, the polarization maps show a hint of an ``hourglass''
perpendicular to the outflow.

\subsection{Comparison of the Magnetic Field Structures in IRAS~16293 with 
IRAS~4A\label{Compara}}

The magnetic field around source A shows the typical ``hourglass''
morphology that is expected from theoretical calculations. As
discussed earlier, this is the second low mass star forming region
where an ``hourglass'' configuration has been observed, the other one
being  IRAS~4A \citep{Girart06}. It is interesting to show  that
while there are some similarities, some significant disparities also exist 
as well:
\begin{itemize}

\item Both sources appear to be part of multiple star systems, with
two dusty main components that have a similar separation: 400~AU and
750~AU for IRAS~4A and IRAS~16293 respectively \citep{Wootten89,
Looney00}.  The total bolometric luminosity, the total mass of the
dense envelope, and the envelope radius surrounding the protostars
are also not too different. IRAS~16293 is only slightly more luminous
but is less massive than IRAS~4A \citep{Sandell91, Correia04} and
powerful outflows originate from both objects.  

\item Another similarity is the contribution of the magnetic energy
to the dynamics of the system. Since both sources appear to be
contracting \citep{diFrancesco01, Chandler05}, their cores must be
approximately supercritical. However the magnetic energy dominates
over the turbulent energy in both of the sources.

\item The biggest difference is that in IRAS~4A the ``hourglass''
morphology is detected in the massive circumbinary envelope,  whereas
in IRAS~16293, the ``hourglass'' is detected in the more compact
circumstellar envelope around source A. In addition, towards
IRAS~16293 most of the emission arises from the two circumstellar
envelopes and from the circumstellar disks (particularly in source
B), with only a small contribution from the circumbinary envelope,
while in IRAS~4A the circumbinary contribution is significant.
In order to quantitatively confirm this, we measured the average
visibility amplitude in the 10--20~k$\lambda$ and 60--75~k$\lambda$
ranges for IRAS~16293, and in the 20--40~k$\lambda$ and 120--150~k$\lambda$
ranges for IRAS~4A (this source is about two times farther than
IRAS~16293, so in order to trace the same physical scale the
visibility range has to be twice as large). The amplitude ratio
between the shortest and longest baselines are $2.1\pm0.2$ and
$1.4\pm0.1$ for sources A and B, whereas it is significantly  larger
for IRAS~4A1 and IRAS~4A2, $4.1\pm0.3$ and $3.2\pm0.2$, respectively.

\item The total integrated fractional polarization from source A
($\simeq 0.5$\%) is much lower than that obtained for IRAS~4A
($\simeq 3.3$\%). One reason for this  could be differences in grain
properties that lead to low alignment efficiencies.  Nevertheless,
we cannot discard the possibility that the more compact emission
detected in IRAS~16293 may also be responsible for this lower
fractional polarization (beam smearing if there is complex unresolved
magnetic structure).

\item A remarkable discrepancy is
the difference in the  orientations of the magnetic field axis and
the outflow axis. For IRAS~4A both axes are not aligned, however
the difference, $\sim 40\arcdeg$ is much less than that in IRAS~16293
where the main active outflow is nearly perpendicular to the main
direction of the field (see \S~\ref{Kine}).  

\end{itemize}

These differences suggest that IRAS~16293 is probably in a more
evolved evolutionary stage and has more mass at smaller scales than
IRAS~4A.  The fraction of the total mass that is in the circumbinary
envelope is larger in IRAS~4A. A significant amount of the mass has
already fallen onto the circumstellar envelopes and the circumstellar
disks of the two main protostellar components of the binary in
IRAS~16293.  The magnetic field configuration also supports this
scenario.  It is also worth noting that the single-dish dust
polarization maps in IRAS~4A \citep{Attard09} trace a uniform
magnetic field which is in agreement with the higher angular resolution SMA
maps of \citet{Girart06}, whereas this is certainly not the case
in IRAS~16293 \citep{Matthews09}.  Tentatively, this could be explained
with IRAS~16293 being a more evolved region,  where the outflow
activity has put significant turbulent energy at the scales traced
by the single-dish maps.

\subsection{The Magnetic Field Morphologies of IRAS~16293 on
Various Sizescales}

IRAS~16293 is located close to the geometrical center of the
L1689-northwest(NW) filament, which also contains the IRAS 16293-2422E
core (which is approximately 2$\arcmin$ to the east). This filament
extends about $10\arcmin$ (0.4~pc) in the NW-SE direction
\citep{Nutter06}. This is the direction that the overall dust
emission appears to have from the SMA maps and is also more apparent
in the BIMA maps of the 2.7~mm dust emission at scales of $20\arcsec$
\citep{Looney00}.  This large scale filament, which could conceivably
be contained in a single magnetic flux tube, appears to break up
into at least two components towards IRAS~16293, sources A and B.
Theoretical calculations by \citet{Mouschovias91} shows that the
process of ambipolar diffusion can initiate single stage fragmentation
along the length  of a flux tube.  The fragmentation, in the case
of such a flux tube, is more likely to occur along its length which
is  parallel to the magnetic field, and is in agreement with our
observations.  Typically not more than three fragments are formed
with each having a mass $\sim 1 M_\odot$.  The fragmentation in
magnetically subcritical clouds occurs when the hydromagnetic waves
decay over length scales smaller than the Alfv\'en length scale and
this can occur when the densities are in the range of $10^3$--$10^6$
cm$^{-3}$.

The SMA dust emission polarization observations conducted by us
only sample the magnetic field on a relatively small scale (angular
size $<15\arcsec$).  Comparable measurements with other instruments
(see \S~\ref{ResultsDustPol}) were also similarly restricted in the
sizescales probed. The magnetic field structure on much larger
scales were traced from observations of polarization due to dust
absorption at optical wavelengths \citep{Vrba76}.  They found that
near the Ophiuchus cloud, the directions of polarization of the
background starlight were approximately uniform.  If so this must
point to a strong and ordered magnetic field. Near-IR polarization
observations, that probe deeper extinctions also show a direction that
is similar to the optical polarization data \citep{Wilking79,
Sato88}.  However, the nearest polarization vectors seen in absorption
are at least $\gtrsim 1\arcdeg$ away from IRAS~16293, which is
roughly 3~pc projected in the plane of the sky and therefore the 
small and the large scale field structures cannot be easily connected.

The ordered magnetic field structure observed by us indicates that
the fields appear to be strong and are not significantly affected
by turbulence on these small scales. However, this may not be the
case on intermediate scales. The SCUBA polarimetry observations of
the IRAS~16293 region \citep{Matthews09}  containing sources A and
B as well as source E show that the polarization position angles
are considerably less ordered. This suggests that turbulence could
possibly dominate on intermediate scales within the molecular cloud
while magnetic fields regulate star formation activity on smaller
scales where collapse signatures are seen in the mapped magnetic
field topology.  However, the sample of polarization vectors detected
is small, and most of the detections have a  low signal to noise
(and consequently have greater position angle errors). Further
higher sensitivity submm single-dish observations are needed in
order to confirm if the dispersion observed is real.

There have been some measurements of the strength of the line of
sight component of the magnetic field (using the Zeeman effect in
HI) in the $\rho$~Ophiuchus cloud by \citet{Goodman94}. While they
were able to measure the field strength significantly at a number
of positions in this region, none of the detections are located in
L1689.  However, it is not clear whether this implies a low magnetic
field strength in the low density gas component of the cloud, or a
combination of projection (field close to the plane of the sky) and
beam smearing (significant structure within the beam gets averaged
out) effects.

\section{Conclusions and Summary}

The installation of a polarimetry system on the SMA has enabled us
to map the magnetic field structure in the ISM, especially in young
star forming regions.  Using this system, we have obtained high
angular resolution and high sensitivity maps of the magnetic field
structure in IRAS~16293 through observations of the polarized dust
continuum. These observations significantly improve on the earlier
measurements which detected extremely low fractional polarization.
Our detections indicate that those early attempts were limited in
their sensitivity and angular resolution.  At the same time, the
net polarization over the entire region mapped by us is quite small
($\sim$ 0.2\%) and this is indeed in agreement with past measurements.
It is also apparent that the polarization fraction appears to
decrease towards the center where the intensity of the source
increases. However, it seems to taper off at sufficiently large
values of the intensity especially in source B. This is possibly
due to the effects of limited resolution towards the center where
the peak of the emission occurs. Further higher resolution observations
will be needed to determine if this relationship continues even at
shorter distances from the center.

The two sources A and B have significantly different magnetic field
morphologies.  In source A, our maps show that the magnetic field
has the pinched ``hourglass'' shape that is expected from theoretical
calculations.  In contrast, the field lines in source B, appear to
be quite uniform. Using the Chandrasekhar-Fermi method to obtain
the  field strength and the continuum dust emission to calculate
the mass, we calculate the mass-to-flux ratio to be $\sim 1$.  This
implies that this object is in (or close to) the supercritical stage
and the magnetic field is no longer able to prevent the collapse.
In addition, the ordered field structures indicate that the magnetic
energy likely exceeds the turbulent energy, but is comparable to
the centrifugal energy. However, there are other reported measurements,
that are not as sensitive, which appear to show  that turbulence
increases on more intermediate sizescales within the cloud.
Nevertheless, the largest scales, which are probed in absorption
polarimetry, once again show fairly ordered magnetic field structures.

The evolutionary stages and the ages of the two sources are likely
to be different and the SMA observations of the magnetic field
structure also appear to confirm this view.  The physical and
chemical properties also differ in the two sources.  In source B,
most of the mass at scales of a few hundred AU has already been
accreted onto a compact, optically thick disk, while source A still
has a significant fraction of its mass in it's circumstellar envelope.
Furthermore, the nature of the outflow activity is quite different
in the two sources: source A shows significant activity and drives
at least two powerful outflows and possibly a third compact outflow
which we have detected in SiO emission, while source B shows no
ongoing activity except for a possible fossil or remnant outflow.
Furthermore, source A appears to have a richer chemical environment
than source B.  It is possible that source B may not be a true
Class~0 protostar and could possibly be a transitional object between
Class~0 and Class~I.  There appears to be a strong misalignment
between the outflow direction in source A and the magnetic field
axis. This is in approximate agreement with theoretical model
predictions when the magnetic energy is not significantly greater
than the centrifugal energy.

This is the second such low-mass star forming region in which it
has been shown that magnetic fields appear to dominate over turbulence.
However, the magnetic field morphology has been mapped in only a
small number of such star forming regions. Improvements in telescope
sensitivity to the polarized flux density will allow us to significantly
expand this sample. With the increase in the number of objects
studied, it will be possible to get a clearer and statistically
significant picture and thereby address the all important question:
Which process plays a dominant role in the star forming process ---
turbulence or magnetic fields?

\acknowledgments
RR wishes to thank all the staff at the SMA in Hawaii, as well as
the instrumentation groups in Cambridge, MA and Taipei for facilitating
the installation and efficient operation of the polarization
instrumentation.  JMG thanks the ASIAA support for the stay in Hilo,
HI.  JMG is supported by Ministerio de Ciencia e Innovaci\'on (Spain)
grant AYA2008-06189-C03.  JMG also acknowledges support from AGAUR
(Generalitat de Catalunya) travel grant 2008-BE2-00203.

\clearpage


\clearpage

\begin{figure}
\includegraphics[scale=0.68]{i16293_PAdist.eps}
\caption{ From bottom to top: Position angle (PA) distribution of
the polarization vectors around source A with respect to the average
value, $<$PA$> = 95.0 \arcdeg$.  PA distribution of the polarization
vectors around source B with respect to the average value, $<$PA$>
= 24.0 \arcdeg$. PA distribution of the residual from the parabolic
fitting done to the source A data and applied to all the data (i.e.
including source B), which is shown in Figure~\ref{Bparabola}.
\label{PAdist}}
\end{figure}


\clearpage

\begin{deluxetable}{lcc}
\tablecaption{Details of the SMA observations \label{obspar}}
\tablehead{\colhead{Date} & \colhead{6$^{\rm th}$ April, 2006} &
\colhead{9$^{\rm th}$ April, 2009}}
\startdata
Opacity\tablenotemark{a} & 0.04 & 0.1\\
Array Configuration & Compact\tablenotemark{b} & Compact\tablenotemark{b} \\
Local Oscillator Frequency (GHz) & 341.5 & 341.5 \\
Bandwidth per sideband (GHz) & $\sim$2.0 & $\sim$2.0\\
Spectral Channels per sideband & 3072 & 3072 \\
Velocity resolution (\kms) & 0.7 & 0.7 \\
\enddata
\tablenotetext{a}{Atmospheric attenuation at 225 GHz from the CSO tau meter}
\tablenotetext{b}{Provides projected baseline lengths from 7 meters
to 70 meters}
\end{deluxetable}


\begin{deluxetable}{lrcrcc}
\tablecaption{Continuum and spectral line parameters\label{tobserva}}
\tablewidth{0pt}
\tablehead{
\multicolumn{2}{c}{}&
\multicolumn{2}{c}{Synthesized Beam}&
\colhead{Spectral} &
\colhead{$rms$}
\\
\cline{3-4}
\colhead{} & 
\colhead{$\nu$} & 
\colhead{HPBW} & 
\colhead{PA} &
\colhead{Resolution} &
\colhead{Noise}
\\
\colhead{Observation} & 
\colhead{(GHz)} & 
\colhead{(arcsec)}&
\colhead{(deg)}& 
\colhead{(\kms)} &
\colhead{$\!\!\!\!$(mJy~beam$^{-1}$)} 
}
\startdata
Continuum       &  341.5        & $3.10\times1.95$ & $-3.7$ & ---
& 4\tablenotemark{a} \\
CO~3--2	& 345.7960  	& $3.06\times1.59$ &$-13.0$	& 2.11 & 250 \\
SiO~8--7	& 347.3307 	& $3.08\times1.63$ & $-13.6$ 	& 1.40 & 340 \\
\htcop\ 4--3 & 346.9985  	& $3.16\times1.84$ & $-7.9$ 	& 0.70 & 420 \\
\enddata
\tablenotetext{a}{Value for Stokes Q and U. The $rms$ noise of the Stokes I 
map is affected by the limited SMA dynamic range, so it is higher, 
14~mJy~beam$^{-1}$}
\end{deluxetable}


\begin{deluxetable}{lcc}
\tablecaption{Properties of sources A and B\label{tbl-props}}
\tablehead{\colhead{} & \colhead{Source A} & \colhead{Source B}} 
\startdata
R. A. (J2000)\tablenotemark{a} &
$16^h 32 ^m 22\fs877$ & $16^h 32 ^m 22\fs621$ \\
Dec. (J2000)\tablenotemark{a} & 
$-24\arcdeg 28\arcmin 36\farcs34$ & $-24\arcdeg 28\arcmin 32\farcs44$\\
Deconvolved Size\tablenotemark{b} &  
$2\farcs29(2) \times 1\farcs28(2)$ &  $0\farcs95(3) \times 0\farcs59(6)$ \\ 
Deconvolved P.A.\tablenotemark{b} & 
	$4\arcdegÊ\pm 1\arcdeg$ &  $85\arcdeg \pm 5\arcdeg$ \\
Peak Intensity (Jy beam$^{-1})$\tablenotemark{a,c} & 
	$4.10\pm0.01$ & $3.89\pm0.01$\\
Flux Density (Jy)\tablenotemark{b,c} & $6.14\pm0.04$ & $4.43\pm0.03$ \\
Flux Density (Jy)\tablenotemark{d,c} & $6.85\pm0.05$ & $4.62\pm0.04$ \\
Equivalent Radius & $4\arcsec$ & $3\arcsec$ \\
Fractional Pol. (\%)\tablenotemark{e} &  0.5$\pm$0.1 &  1.0$\pm$0.1 \\
Pol. P.A.($\arcdeg$)\tablenotemark{e} & 104$\pm$6 & 26$\pm$3\\
Mass (M$_\odot$) & 0.33 & 0.22\\
Volume Density (cm$^{-3}$) & $4.9\times10^7$ & $7.9\times10^7$\\
Column Density (cm$^{-2}$) & $5.9\times10^{23}$ & $7.1\times10^{23}$\\
\enddata
\tablenotetext{a}{Estimated using Miriad's ``MAXFIT'' task.}
\tablenotetext{b}{Estimated from 2-dimensional Gaussian fittings
to the image using AIPS's ``IMFIT'' task with  a 0.5~Jy~beam$^{-1}$
cutoff to avoid the contribution from the weak extended component.
Number in parenthesis give the uncertainty of the last decimal.}
\tablenotetext{c}{The absolute flux scale is accurate only upto 5\%.}
\tablenotetext{d}{Estimated integrating the flux in the region used
for polarization analysis, i.e. with a 0.15~Jy~beam$^{-1}$ cutoff.}
\tablenotetext{e}{The fractional polarization and position angle
were determined by integrating Stokes I, Q, and U over each source.}
\end{deluxetable}


\begin{deluxetable}{cccccc}
\tablecaption{Table of polarizations in IRAS~16293\label{tbl-listpol}}
\tablehead{ \colhead{RA\tablenotemark{a}} & \colhead{Dec\tablenotemark{a}}
& \colhead{I (Jy beam$^{-1}$) \tablenotemark{b}} & \colhead{P (mJy
beam$^{-1}$)\tablenotemark{c}} & \colhead{p (\%)} & \colhead{$\theta
$ ($ \arcdeg$)}}
\startdata
   1.000 &   -4.500 &    0.342 &   10 &    3.0$\pm$1.3 &    -51.1$\pm$12.5 \\ 
   0.000 &   -3.000 &    1.218 &   12 &    1.0$\pm$0.4 &    -75.0$\pm$11.0 \\ 
  -1.000 &   -3.000 &    0.894 &   14 &    1.6$\pm$0.5 &    -49.1$\pm$9.2 \\ 
   0.000 &   -1.500 &    3.103 &   16 &    0.5$\pm$0.1 &    -86.6$\pm$7.9 \\ 
  -1.000 &   -1.500 &    2.402 &   18 &    0.8$\pm$0.2 &    -63.0$\pm$7.2 \\ 
   0.000 &    0.000 &    3.879 &   17 &    0.4$\pm$0.1 &    -88.2$\pm$7.8 \\ 
  -1.000 &    0.000 &    3.102 &   19 &    0.6$\pm$0.1 &     86.4$\pm$6.8 \\ 
  -2.000 &    0.000 &    0.981 &   10 &    1.1$\pm$0.5 &     68.2$\pm$12.6 \\ 
   2.000 &    1.500 &    0.192 &    8 &    4.2$\pm$2.4 &    -40.6$\pm$16.0 \\ 
   1.000 &    1.500 &    1.014 &   18 &    1.8$\pm$0.4 &    -54.4$\pm$7.1 \\ 
   0.000 &    1.500 &    2.180 &   16 &    0.7$\pm$0.2 &    -84.0$\pm$8.0 \\ 
  -1.000 &    1.500 &    1.893 &   15 &    0.8$\pm$0.2 &     60.6$\pm$8.8 \\ 
  -2.000 &    1.500 &    0.941 &   19 &    2.1$\pm$0.5 &     48.4$\pm$6.7 \\ 
  -3.000 &    1.500 &    1.069 &   15 &    1.4$\pm$0.4 &     45.3$\pm$8.9 \\ 
  -4.000 &    1.500 &    1.206 &   15 &    1.2$\pm$0.4 &     23.3$\pm$8.8 \\ 
  -5.000 &    1.500 &    0.413 &   18 &    4.4$\pm$1.1 &     23.4$\pm$7.2 \\ 
   1.000 &    3.000 &    0.357 &   10 &    2.7$\pm$1.3 &    -58.2$\pm$13.5 \\ 
   0.000 &    3.000 &    0.742 &    9 &    1.2$\pm$0.6 &    -84.4$\pm$14.4 \\ 
  -1.000 &    3.000 &    0.864 &    8 &    0.9$\pm$0.5 &     56.0$\pm$16.5 \\ 
  -2.000 &    3.000 &    1.089 &   14 &    1.3$\pm$0.4 &     35.9$\pm$9.4 \\ 
  -3.000 &    3.000 &    2.577 &   29 &    1.1$\pm$0.2 &     30.5$\pm$4.5 \\ 
  -4.000 &    3.000 &    3.459 &   36 &    1.0$\pm$0.1 &     24.0$\pm$3.6 \\ 
  -5.000 &    3.000 &    1.306 &   25 &    1.9$\pm$0.3 &     14.1$\pm$5.3 \\ 
  -3.000 &    4.500 &    2.089 &   19 &    0.9$\pm$0.2 &     31.5$\pm$7.0 \\ 
  -4.000 &    4.500 &    3.053 &   28 &    0.9$\pm$0.1 &     30.9$\pm$4.6 \\ 
  -5.000 &    4.500 &    1.259 &   17 &    1.4$\pm$0.4 &      9.3$\pm$7.5 \\ 
  -5.000 &    6.000 &    0.369 &   11 &    2.9$\pm$1.2 &    -20.1$\pm$12.3 \\ 
\enddata
\tablenotetext{a}{Offsets in arcseconds from the central pixel
located at position with coordinates of RA=$16^h32^m22\fs9$, and
Dec=$-24\arcdeg28\arcmin36\farcs00$}
\tablenotetext{b}{Values of the Stokes I continuum flux density.
The $rms$ of the Stokes I continuum flux density map is $\sim$17
mJy beam$^{-1}$}
\tablenotetext{c}{Values of the polarized flux density. The $rms$
of the polarized flux density map is $\sim$4 mJy beam$^{-1}$}
\end{deluxetable}


\begin{deluxetable}{cccccc}
\tablecaption{Comparison  of polarizations at different 
wavelengths\label{compols}}
\tablehead{\colhead{Wavelength(mm)} & \colhead{Beamsize ($\arcsec$)}&
\colhead{p (\%)} & \colhead{$\theta$ ($\arcdeg$)} & \colhead{Reference}}
\startdata
2.84 & 5.3$\times$3.1 & 0.7$\pm$0.5 & 30$\pm$7  & Akeson et al. (1997)\\
1.1 & 19 & 2.2$\pm$0.4 & 135$\pm$5 & Tamura et al. (1995)\\
0.88 & 3.2$\times$1.9 & 0.17$\pm$0.01 & 53$\pm$3 & This work\tablenotemark{a} \\
0.8 & 11 & 1.4$\pm$0.5 & 62$\pm$11 & Flett \& Murray (1991)
\enddata
\tablenotetext{a}{The polarization measurement is done by summing
Stokes I, Q, and U in the entire map and then obtaining the polarized
flux density, fractional polarization and position angles. }
\end{deluxetable}

\end{document}